\documentstyle[12pt,epsfig]{article}
\begin{document}
\vspace*{2cm}
\begin{center}
{\Large {   Determination of Azimuthal Anisotropy  of $\pi^0$ from the
Measured Anisotropy of Photons in Ultra-Relativistic Nuclear Collisions }}

\end{center}

\bigskip
\begin{center}

{\small{
\noindent

Rashmi~Raniwala$^{1}$,
Sudhir~Raniwala$^{1}$
and Yogendra~Pathak~Viyogi$^{2}$

}}
\end{center}
\begin{center}
{\small{

{$^{1}$~Physics Department, University of Rajasthan, Jaipur 302004, India} \\
{$^{2}$~Variable Energy Cyclotron Centre,  Calcutta 700064,
  India}
}}           
\end{center}

\begin{abstract} 

A method is suggested to deduce the anisotropy in neutral pions by measuring 
the azimuthal anisotropy of photons  in ultra-relativistic nuclear
collisions.
The ratio of the estimated anisotropy in photons 
to the anisotropy in neutral pions
is seen to scale with a parameter which depends on  
photon multiplicity and anisotropy. This parameter can be
determined from experimental data.

\end{abstract}

\normalsize
\section{Introduction}

In ultra-relativistic nuclear collisions the colliding 
nuclear matter emerges with a pattern that has its 
origin in the  incompressibility of nuclear matter and
the mean field effects.
The observation of collective behaviour in the system 
        validates the use of a hydrodynamic description of the evolution of the collision.
        Within the hydrodynamic model the collective motion results from 
        pressure gradient in the matter, which depends upon the  
        compressibility of the underlying equation of
        state \cite{ollie}. In the case of a phase transition from hadronic
        matter to quark
        gluon plasma, increased number of degrees of freedom may 
lead to a softening of the equation of state 
        \cite{ritter,daniel,rischke,ollieqm}. Collective flow
        manifests  in altered event shapes which can 
        be studied by measuring the momentum distributions of produced 
        particles. The azimuthal distribution of these  event shapes
        are studied using coefficients in their Fourier expansion. 
        One of the major current aims of the study of ultra-relativistic 
        heavy ion experiments is to determine the magnitude of anisotropy
        present in the final state momentum distributions of various particle 
        species, in different regions of phase space. The systematic 
        dependence of this anisotropy on energy, 
        centrality and other collision 
parameters is predicted to be sensitive to the equation of 
        state \cite{ritter}.

     Most of the experimental studies on azimuthal anisotropy have been
        restricted to
     charged particles (a large number of results for charged particle
     measurements can be found in the review article 
        \cite{ritter}). The measured anisotropy of charged particles
  includes effect 
        due to final state Coulomb interactions and  may inhibit a direct
        comparison of results with the predictions of
 strong interaction physics.

       Measurements of anisotropy in the
        azimuthal distribution of photons have been reported recently 
\cite{wa93_flow,wa98_flow}. Since almost 90\% of photons produced in 
ultra-relativistic nuclear collisions originate from the decay of $\pi^0$, the
anisotropy measured in photons should have their origin in the anisotropy
of $\pi^0$ produced in the reaction. 
However, the measured anisotropy in photons will also include some effect
due to decay. The decay introduces non-flow correlations amongst kinematic 
variables of photons due to four-momentum conservation and also 
dilutes the existing correlation in azimuthal angles.
Determination of the effect of decay
should enable deduction of the anisotropy in $\pi^0$.
This would complement the study
of flow using charged particles and would be free from Coulomb effects.
        Such a determination will provide much needed additional information
        to characterise the final state momentum 
distributions \cite{alice_pmd}. A comparison of the anisotropies 
deduced from charged and neutral pions
        would help in estimating the effect of final state
        Coulomb interaction on the charged particles.

In the present work we study the effect of $\pi^0$ decay on the
observed anisotropy of photons for various assumed flow parameters and
multiplicities using a fast simulation technique. Preliminary results
can be found in \cite{alice_pmd,star_pmd,alice-note}. We  
show how the anisotropy parameter for the 
        parent $\pi^0$ distribution
 can be deduced using the measured anisotropy 
        in the azimuthal distribution of photons.

\section{Simulation}

        A complete 
        determination of the flow parameters in different regions 
would require
        a measurement of the three-dimensional momentum distribution of
        all the particles. 
       Several experiments  determine 
        the anisotropy parameters in azimuthal angle 
        distributions (\(dN/d\phi \)) 
        of produced particles.
\cite{wa93_flow,wa98_flow,ceres,na49}. In the present work
        we restrict ourselves to the study of photon multiplicity in a 
 detector having full azimuthal acceptance 
in one
unit of pseudo-rapidity suitably chosen with practical detectors in mind
\cite{wa98_flow,alice_pmd,star_pmd}. 
We generate events having azimuthally 
anisotropic shapes and analyse them using methods 
employed for the analysis of experimental data.

\subsection{Anisotropy}

        Let us assume that the event shape in the azimuthal plane can 
        be described by an arbitrary ellipse. The most general equation 
        of an ellipse in polar coordinates is
\begin{eqnarray}
      r(\phi)= \frac{1}{2\pi}\left[1 + 2v_1 \cos(\phi -\psi_1) +
 2v_2 \cos 2(\phi-\psi_2)\right]
\label{aniso}
\end{eqnarray}

\noindent where $\psi_1$ and $\psi_2$ define the direction of the shift 
        of the centroid of the ellipse from the origin and the orientation of 
        the major axis of the ellipse respectively. $v_1$ is  
        the magnitude of the shift of the centroid and $v_2$ is a measure
        of the difference in 
        the major and the minor axes of the ellipse. 
 $v_1$ and  $v_2$ denote the first order and the second 
        order flow components, termed as directed and elliptic flow.

\subsection{Event generation} 

        For the fast simulation, a set of events consisting of 
$\pi^+,\pi^-$ and $\pi^0$ is generated. The particle multiplicity 
in the event is selected in 
such a way that the multiplicities of charged particles ($\pi^+,\pi^-$) and
decay photons within the one unit of pseudo-rapidity acceptance of the given
detector are in the range of 100 to 4000.
Such a wide range of multiplicity
corresponds to varying centralities encompassing  the SPS, RHIC and LHC 
energies and  
serves to determine the effect of 
 fluctuations due to finite number of particles.
The multiplicity distribution of particles  is assumed to be Gaussian with a
 width $\sigma = 2\sqrt{M}$ for  a mean multiplicity M.

        We assume the azimuthal distributions to have initial flow components
$v_m^{in}$, where $m$ (=1,2) denotes the order of flow. Using these values, 
the azimuthal angles of the charged and neutral 
pions  are assigned according to the distribution given by 
eqn.~(\ref{aniso}). 
$\psi_1$ is chosen    randomly. For
        the present work we have chosen $\psi_2$ = $\psi_1$, 
which corresponds to the major axis of the ellipse along the direction
of the shift of the centroid. Physically this represents
         in-plane flow.
The data are also
        generated for different values of $v_1^{in}$ and $v_2^{in}$, 
corresponding to different initial flow.

        The pseudo-rapidity and $p_T$-distributions are taken from a
parametrised form of the HIJING event generator \cite{hijing}
at the relevant energy.
        Neutral pions are generated in the  pseudo-rapidity
        interval extending to 1.5 units on both
        sides of the assumed acceptance of the detector. These are allowed
        to decay and the $\eta$, $p_T$ and $\phi$ 
        of the photons are obtained. The photons falling within the 
acceptance region of the detector are used for further analysis. 

\subsection{Analysis}
The set of events thus generated is 
        analysed to extract the anisotropy parameters $v_m$ using the 
Fourier expansion technique in a  manner identical to the
analysis of experimental data.
         The present analysis is based on 10000 events for each 
combination of multiplicity and anisotropy.

        The generated data have been analysed to obtain the following:
\begin{enumerate}

 \item $\psi_1^{est}$ : the estimated direction of  the first order  
        event plane. Because of finite particle multiplicity the estimated 
        event plane $\psi_1^{est}$ in the generated data fluctuates about 
        the actual event plane, $\psi_1$, with a spread that depends on 
the initial anisotropy and the multiplicity.  
        \item $v_1'$ = $\langle \cos (\phi - \psi_1^{est}) \rangle $ :  a
measure of the directed flow in the direction of $\psi^{est}_1$, 
where the average is over
        all particles of all events. 

        \item $\psi_2^{est}$ : the estimated direction of 
second order event plane.
 This also fluctuates 
        about the actual direction $\psi_2$. 
        \item $v_2'$ = $\langle \cos 2(\phi - \psi_2^{est}) \rangle $ : a 
        measure of the ellipticity about $\psi_2^{est}$
where the average is over all particles of all the events. This is a measure
of
the difference of the major axis and 
        the minor axis of the event shape.
 \end{enumerate}
 
       Since the estimated event plane 
        differs from the actual event plane,  the values of $v_m'$ are 
        systematically lower by a factor 
        $\langle \cos ~m (\psi_m - \psi_m^{est}) \rangle$, 
where the average is 
        over all events for a particular sample. This factor is the 
        resolution correction factor (RCF). The values $v_m$ that
represent the magnitude of directed ($m$ =1) and elliptic 
($m$ = 2) flow, for the sample of events, are 
obtained by $v_m'$/RCF. 

        Experimentally, RCF is obtained using the  subevent method
        described in \cite{posk}. 
         Here every event is  divided randomly into two 
         subevents  of equal multiplicity and the angle $\psi_m$ 
is determined for each subevent.
This enables a determination 
        of a parameter $\chi_m$ directly from the experimental data
using the relation \cite{posk,olli1}:
\begin{eqnarray}
        \frac{N_{events} ( m | {\psi_m^a - \psi_m^b}| > \pi/2)}{N_{total}} 
= \frac{e^{-\frac{\chi_m^2}{4}}}{2}
\label{chim}
\end{eqnarray}
\noindent where $N_{total}$  denotes the total number of
events, $\psi_m^a$, $\psi_m^b$ are the estimated 
angles of the two subevents (labelled
$a$ and $b$) and the numerator on the left denotes the
 number of events having the angle between subevents greater
than $\pi/2m$. 
The parameter $\chi_m$ 
  is then used to determine RCF according to the relation given in the Ref.
\cite{posk}.

        The generated azimuthal distribution of the charged pions are 
        also analysed in a manner similar to the azimuthal distribution
        of photons. Comparing the results of charged particles and photons
        enables a determination of the effect of decay. 

\begin{figure}[h] 
\centerline{\epsfig{file=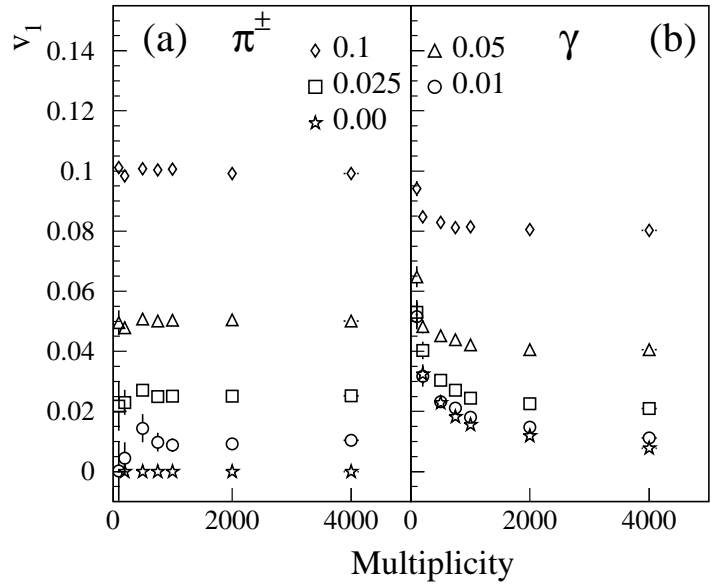,width=12cm}}
\caption{ Estimated value  $v_1$ of directed flow for charged particles
 and photons as a function 
of multiplicity. Symbols denote different initial anisotropies 
varying from 0 to 0.1. 
For photons the
initial anisotropy  was introduced in the $\pi^0$ distribution.
The error bars are statistical.}
\label{v1mult}
\end{figure}

\section{Results }

\subsection{Effect of decay}

\begin{figure}[h] 
\centerline{\epsfig{file=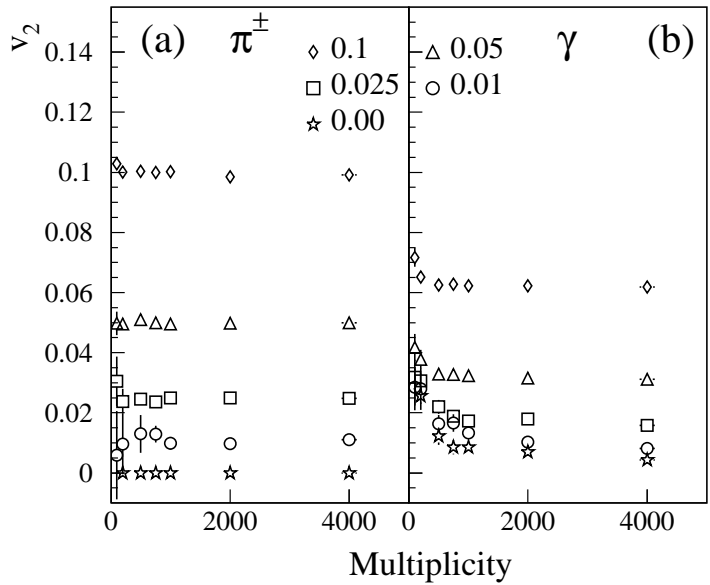,width=12cm}}
\caption{ Estimated value  $v_2$ of elliptic flow for charged particles
 and photons as a function 
of multiplicity. Symbols denote different initial anisotropies varying from
0 to 0.1. 
For photons the
initial anisotropy  was introduced in the $\pi^0$-distribution.
The error bars are statistical.}
\label{v2mult}
\end{figure}

The results for the estimated anisotropy $v_1$ 
for directed flow  are shown
        in Fig.~\ref{v1mult}(a) as a function of multiplicity for charged 
particles and Fig.~\ref{v1mult}(b) for photons for varying values of
$v_1^{in}$.
Similar results for elliptic flow are shown in Fig.~\ref{v2mult}. 
For charged particles the analysis technique
correctly reproduces 
the initial 
anisotropy for all multiplicities for both directed and elliptic flow, 
demonstrating the validity of the
technique used.
For lower multiplicities the errors on the estimated values
are larger due to greater fluctuations.

The same analysis technique, when applied to photons, yields values of
 $v_m$ that 
are generally different from the initial anisotropy present in the
$\pi^0$ distribution. 
        This difference in the anisotropy values for photons and $\pi^0$ 
is attributed 
        to two effects of decay :  (i) it dilutes the initial flow by 
        randomising the existing azimuthal correlation, and (ii) it mimics 
        flow, the correlations  due to momentum conservation 
        producing  an anisotropy in the final state distributions.

To estimate the effect of momentum conservation correlation
we generated azimuthally symmetric distributions, corresponding to 
$v_m^{in}$=0. 
The decay photons falling within the acceptance of 
the detector were analysed, and the values of $v_m$ obtained for 
them are also shown in Fig.~\ref{v1mult}(b) and in Fig.~\ref{v2mult}(b). 
One observes that the deduced anisotropy in photons is always non-zero.
At low multiplicities the values can be as high as those observed in charged
particle measurements \cite{na49}.
The values for elliptic flow are systematically lower than for 
directed flow, implying lesser effect of momentum conservation
correlation on the elliptic event shape.

For all values of initial anisotropy, the estimated anisotropy in
photons decreases with increasing multiplicity. 
At low multiplicity the correlation due to momentum conservation competes
with dilution due to decay.
The relative importance depends on the initial anisotropy in neutral pions.
For low anisotropy in $\pi^0$, this leads 
to enhanced anisotropy in photons.

The above results hold for both orders of flow.
It is  seen that the decay dilutes
the elliptic flow more than it dilutes directed flow.

\subsection{Estimation of $\pi^0$ flow}

\begin{figure}[h]
\vspace*{-0.8cm} 
\centerline{\epsfig{file=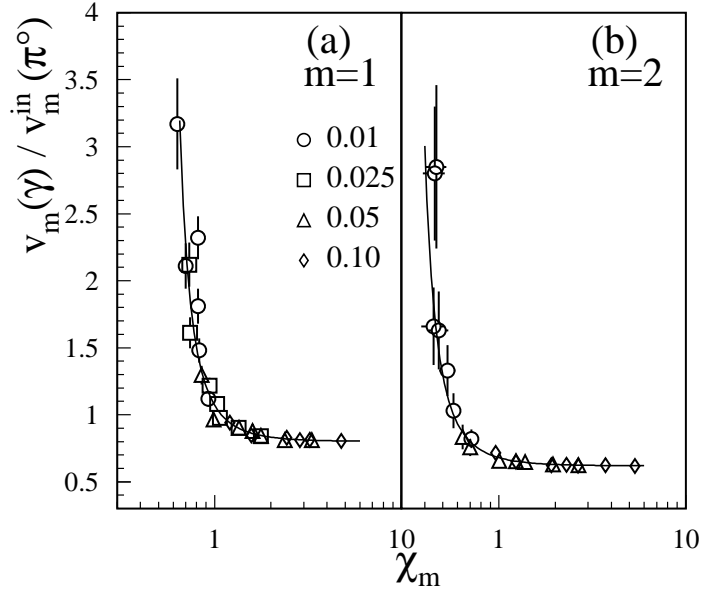,width=12cm}}
\vspace*{-0.3cm} 
\caption{ Ratio of estimated value of anisotropy in photons to the initial
 value in $\pi^0$ 
 as a function of measured anisotropy parameter $\chi$,
 for different values of anisotropy and multiplicity (a) for $m$ = 1, 
directed flow and (b) for $ m$ = 2, elliptic flow. 
Symbols denote different initial $\pi^0$ anisotropies.
Logarithmic $x$-axis is used for better visualisation of the results at lower
$\chi_m$ values.
The error bars are statistical.}
\label{flow_scale}
\end{figure}

 While the effect of decay is seen to depend on the values of 
multiplicity and
        the amount of initial flow, the ratio of anisotropy observed
        in photon distributions,
$v_m(\gamma)$, to that in the parent $\pi^0$ 
distribution, $v_m^{in}(\pi^0)$, is seen to 
        scale with the parameter $\chi_m$. Fig. 
        ~\ref{flow_scale} shows this ratio  
        as a function of $\chi_m$ for a
        large combination of values of anisotropy and multiplicity.

\noindent We have parametrised the observed scaling behaviour by a relation :
\begin{eqnarray}
        \frac{v (\gamma)}{v^{in}(\pi^0)} = \frac{a}{(\chi-b)^2} + c
\end{eqnarray}

\noindent The relation was motivated by the fact that the effect of 
two particle correlations scales as 1/M \cite{posk} 
and $\chi_m$ is proportional to $\sqrt{M}$.
If the exponent is chosen a free parameter, the best-fit value  
 varies between 
1.9 and 2.1, depending upon the number of points used in
the fit.

The continuous curves in Fig.~\ref{flow_scale} show the fitted functions for
both directed and elliptic flow.
        The values of the constants for different values of $m$ are given
        in Table~\ref{fit}.  
The values of $b$ reflect the lower limit of $\chi$ upto which the
anisotropy in $\pi^0$ can be deduced by this method. The lower value of 
$b$ for $m$=2 implies that elliptic flow in neutral pions can be studied
for lower anisotropy and multiplicity. 
 $c$ is the limiting value of the fractional
anisotropy remaining in photons.

\begin{table}[h]
\begin{center}
\caption{Fitted values of the constants of the scaling relation.}
\label{fit}
\vskip 3mm
\begin{tabular} {|c|c|c|c|} \hline
Order ($m$) & $a$ & $b$ & $c$ \\ \hline
1 & 0.075 & 0.473  & 0.801 \\
2 & 0.031 & 0.286 & 0.619 \\ \hline
\end{tabular}
\end{center}
\end{table}

It is noteworthy that a simple parameterisation
can be used up to quite low values of $\chi$. 
The parameter $\chi_m$ depends on anisotropy and multiplicity
($\chi_m = v_m \sqrt{2M}$) and can be determined from experimental data using
eqn.~(\ref{chim}). Combining this with the  anisotropy values $v_m$
determined from the experimental data on photon distributions, 
one can deduce the anisotropy
present in the parent $\pi^0$ distribution using the above scaling relation.

\begin{figure}[h] 
\vspace*{-1cm}
\centerline{\epsfig{file=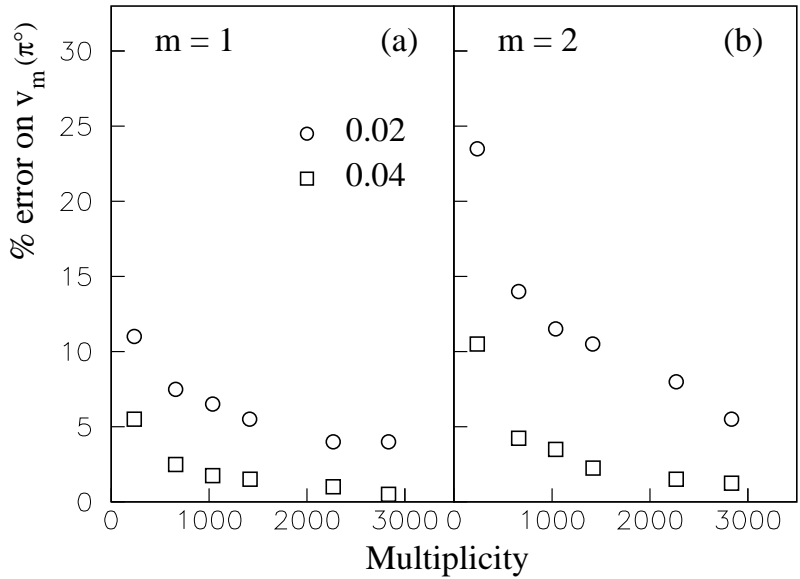,width=12cm}}
\caption{ 
The percentage error on the value of $\pi^0$ anisotropy, deduced 
using the scaling relation, as a function of multiplicity of photons for
(a) directed flow and (b) elliptic flow. The circles represent 
 initial anisotropy $v_m(\pi^0)$= 0.02 and squares represent
$v_m(\pi^0)$= 0.04.}
\label{flow_error}
\end{figure}

To check the efficacy of the method, we simulated another data
set with $v_m^{in}$ = 0.02 and $v_m^{in}$ = 0.04 and for a range 
of 
multiplicities varying between 250 and 3000. 
This data set was analysed as earlier and 
the value of anisotropy in 
neutral pions was estimated 
using the scaling
relation.
The estimated values are close to the input values, the ratio of the two
 deviating
 from unity only for the cases
 when both the multiplicity and the initial flow
are small.

The error on the deduced $\pi^0$ anisotropy will be small for large values of 
$\chi$, as is evident from the local slope of the curve in
 Fig.~\ref{flow_scale}. 
The percentage errors on the deduced anisotropy values are shown as 
a function of multiplicity in  Fig.~\ref{flow_error}
 for both  orders of flow. 
These have been calculated by taking the statistical errors in
$v_m(\gamma)$ and $\chi_m$, both experimentally measurable quantities.
The errors are generally within 10\%, except for elliptic flow at low 
multiplicities.


\section{Summary}
A set of events having built-in azimuthal anisotropies has been
analysed using techniques employed for the analysis of experimental data.
The anisotropies are introduced in both charged and neutral pion 
distributions. Photon distributions are obtained from the $\pi^0$ 
distributions.  The pion distributions are generated to have the 
desired multiplicities in the range 100 to 4000  in one unit 
of pseudo-rapidity. 

Analysis of charged particle distribution leads to estimates 
of anisotropy values in close agreement with the initial anisotropy present
in the event. Due to the effect of decay, the values obtained from 
photon distributions are
different, being lower, or higher, than the initial anisotropy depending
on the value of the initial anisotropy and multiplicity.

The ratio of estimated anisotropy in photon distributions to the initial
anisotropy in $\pi^0$ distributions is seen to scale with a parameter
$\chi$ which can be determined from experimental data and depends on
both anisotropy and multiplicity. This scaling behaviour provides us with a 
simple relation to deduce the initial anisotropy in $\pi^0$ distributions
by the experimental study of photon distributions. The anisotropy is 
obtained with varying precision which improves when either flow or
multiplicity increases.

\vskip 1cm
\noindent{\bf Acknowledgements}

We thank Drs. Sardar Singh and Tapan Nayak for useful discussions and Dr.
Dinesh Srivastava for a critical reading of the manuscript. Financial
support from the Department of Science and Technology, Government of India,
is gratefully acknowledged.

\clearpage
\end{document}